\documentclass[twocolumn,prl]{revtex4}

\usepackage{graphicx}
\usepackage{dcolumn}
\usepackage{bm}
 
\def\bea{\begin{eqnarray}}
\def\eea{\end{eqnarray}}

\def\ben{\begin{equation}}
\def\een{\end{equation}}

\input rotate

\begin{document}

\preprint{DIPC}

\title{Hellman-Feynman operator sampling in Diffusion Monte Carlo
calculations}
\author{R. Gaudoin$^1$ and J. M. Pitarke$^{2,3}$}
\affiliation{
$^1$Donostia International Physics Center (DIPC),
Manuel de Lardizabal Pasealekua, E-20018 Donostia, Basque Country, Spain\\
$^2$CIC nanoGUNE Consolider, Mikeletegi Pasealekua 56, E-2009 Donostia, Basque
Country, Spain\\
$^2$Materia Kondentsatuaren Fisika Saila, UPV/EHU, and Unidad F\'\i
sica Materiales
CSIC-UPV/EHU, 644 Posta kutxatila, E-48080
Bilbo, Basque Country, Spain}

\date{\today}

\begin{abstract}
Diffusion Monte Carlo (DMC) calculations typically yield highly accurate
results in solid-state and quantum-chemical calculations. However, operators
that do not
commute with the Hamiltonian are at best sampled correctly up to second order
in the
error of the underlying trial wavefunction, once simple corrections have been
applied. 
This error is of the same order as that for the energy in variational
calculations. Operators
that suffer from these problems include potential energies and the
density. This paper presents a new method, based on the Hellman-Feynman
theorem, 
for the correct DMC sampling of
all operators diagonal in real space. Our method is easy to implement in
any standard DMC code.
\end{abstract}

\maketitle

Diffusion Monte Carlo (DMC) is widely used for the computation of
properties of solids and molecules~\cite{qmc}.  
Frequently, it is used as a check on other methods~\cite{check} or even as an
input~\cite{input}. It is 
therefore very important that DMC be as accurate as possible. 
However, other than for the total energy, standard  DMC calculatioins are not
as definitive as one would hope, since operators that do not commute with the
Hamiltonian cannot be sampled exactly within standard DMC. Here we present a
simple yet effective addition
to standard DMC that plugs that gap and is easy to implement.
 
DMC by construction yields the normalized
expectation value $\langle \hat{O}\rangle_{DMC}=
 \langle \Psi_T| \hat{O}|\Psi^{fn}_0 
\rangle/\langle \Psi_T|\Psi^{fn}_0\rangle$,  
which is generally not the true ground-state expectation value
$\langle \hat{O}\rangle=
 \langle \Psi_0| \hat{O}|\Psi_0 
\rangle/\langle \Psi_0|\Psi_0\rangle$. In fact, it is not even 
$\langle \hat{O}\rangle_{fn}=
 \langle \Psi^{fn}_0| \hat{O}|\Psi^{fn}_0 
\rangle/\langle \Psi^{fn}_0|\Psi^{fn}_0\rangle$, the ground-state expectation
value 
constrained by a nodal structure of the fermionic many-body wavefunction that
is given by $\Psi_T$. 
$\Psi_T$ is a trial wavefunction that approximates the generally unknown
ground-state
wavefunction $\Psi_0$ and is real.  In its basic and most common form
$\Psi^{fn}_0$ is the ground state for a fixed nodal structure given by that
of $\Psi_T$. In addition to this fixed-node
approximation operators that do not commute with the Hamiltonian are
generally subject to a further error, the leading term of which is linear in
the
difference between $\Psi_T$ and $\Psi^{fn}_0$. In conjunction with Variational 
Monte Carlo (VMC), this error can be reduced by one order~\cite{cep_err} by
using $\langle \hat{O}\rangle_{cDMC}=2\langle \hat{O}\rangle_{DMC}-\langle
\hat{O}\rangle_{VMC}$.
Correct sampling can be achieved, e.g. by 
using forward walking~\cite{fwalking}, reptation Monte Carlo~\cite{rmc}, and 
other methods~\cite{no_tagging}.
Many of these methods aim to sample $\Psi_0^{fn}\Psi_0^{fn}$, rather than the
usual DMC distribution 
$\Psi_T\Psi_0^{fn}$. They  are therefore not straight forward additions to the
DMC algorithm. Alternatively, the Virial theorem or the related
Hellman-Feynman (HF) theorem~\cite{hf} can be used to evaluate operator
expectation values~\cite{ob} which in the case
of DMC, however, involves numerical derivatives of noisy data. 

In this Letter, we present a method based on the application of the HF theorem
to the
DMC algorithm directly. Our method -  Hellman-Feynman sampling (HFS) - can be 
tagged onto the usual sampling of
operators with nearly no extra computational overhead. The aim is to maintain
the basic DMC 
algorithm that samples $\Psi_T\Psi_0^{fn}$. 
Now, the total energy is evaluated correctly within standard DMC, and
crucially operator expectation values can be cast as HF derivatives of the
total
energy.
Keeping in mind that ultimately the DMC algorithm is nothing but a large
sum that yields the total energy, we see the HF derivative can be applied
without problem
to the algorithm itself!
One advantage over numerical derivatives is that the resulting formula can
handle
several operators simultaneously in a single DMC run and maintaining 
orbital occupancy for perturbed Hamiltonians 
ceases to be a problem.
 The DMC algorithm only involves numbers, so 
non-commutability of operators - the source of the difficulties -
is no issue either. Writing down the DMC algorithm as a mathematical formula
and applying the
HF derivative to it yields an object that when sampled using standard DMC
produces
the exact operator expectation value. It has to by construction. 

In the following, we present a schematic overview of the DMC algorithm,
which however is sufficient to derive the relevant formulas.
The basic idea of DMC is to split the 
imaginary-time propagator
$\exp(-\Delta t \hat{H}) \approx \exp(-\Delta t \hat{T})\exp(-\Delta t
\hat{V})$ 
for sufficiently small time intervals $\Delta t$ 
into a kinetic and potential term
and then to iterate it. This ultimately~\cite{note_dmc_alg} gives rise to a 
 real-space drift-diffusion process sampled using Monte Carlo (MC), augmented 
by an exponential growth term whereby $N_w$ so-called walkers 
are propagated in parallel. Courtesy of this growth term, at each propagation
or (imaginary) time
step $i$ the walker $j$ acquires a
multiplicative weight: $e^{-\Delta t \left(E^L_{i,j}-\tilde{E}^0_i\right)}$, where
$E^L_{i,j}=\hat{H}\Psi_T/\Psi_T$ evaluated at the real-space position of walker
$j$ at time step $i$ and $\tilde{E}^0_i$ is an estimate for the ground-state energy
also at time step $i$. 
The total weight of walker $j$ at time step $i$ becomes
\ben
\label{def_omega}
\omega_{i,j}=\prod_{k=1}^{i}
e^{-\Delta t \left(E^L_{k,j}-E^0_{i}\right)},\, \mbox{where}\,
E^0_i=\frac{1}{i}\sum_{l=1}^{i} \tilde{E}^0_l
\een
and the presence of $E^0_i$ ensures normalization.
At time step $i$ the estimator for an operator that a DMC code yields is 
\ben
\label{dmc_alg}
\overline{O^L_i} = \sum_j^{N_w} \omega_{i,j} O_{i,j}^L,
\een
where
$O_{i,j}^L=\hat{O}\Psi_T/\Psi_T$ and the wavefunction $\Psi_T$ is evaluated 
for walker $j$ at time step $i$. For brevity, we use this bar-average
$\overline{O^L_i}$ 
where applicable and note that $\overline{O^L_i}$ has to be averaged over all
$i$
to yield the final DMC estimate $\langle \hat{O}\rangle_{DMC}$. Since  the
ground-state energy is not
known, an estimate chosen such that Eq. (\ref{def_omega}) remains normalized
has to be used. This is the growth 
estimator  $E^0_i$~\cite{Etilde} and
is updated at each step, hence the index $i$. 
Note that $E^0_i$ is independent of $j$, i.e. it is the same for every walker
and thus a property of the DMC process as a whole.
For reasons of numerical stability, DMC is implemented by allowing walkers to
die 
or multiply such that the walker's survival probability optionally augmented by 
residual weights corresponds to Eq. (\ref{def_omega}).  

Given a perturbed Hamiltonian $\hat{H}(\alpha)=\hat{H}+\alpha \hat{O}$ and
the associated fixed-node ground state energy 
$ E_{fn}(\alpha)=\langle\hat{H}\rangle_{DMC}$, 
first-order perturbation theory for $\Psi_0^{fn}$ yields a fixed node
equivalent of the HF theorem~\cite{HFproblem}
\ben 
\label{hf_exp}
\langle O \rangle_{fn} =\left.
\frac{\partial E_{fn}(\alpha)}{\partial\alpha} 
\right|_{\alpha=0},
\een 
where  $\langle O \rangle_{fn}$ converges to the correct 
ground-state  expectation value as the nodes of $\Psi_T$ become 
exact though $\Psi_T$ itself need not. Note that while 
$\langle\hat{H}\rangle_{DMC}=\langle\hat{H}\rangle_{fn}$ we have
$\langle\hat{O}\rangle_{DMC}\neq\langle\hat{O}\rangle_{fn}$,
unless $[\hat{O},\hat{H}]=0$, so Eq. (\ref{hf_exp}) is not trivial. 
The energy $E_{fn}(\alpha)$ is
accessible exactly within standard DMC as the Hamiltonian
$\hat{H}(\alpha)$
commutes with itself. 
Analytic operator estimators can then be derived by
applying the HF theorem to the formula expressing the DMC algorithm Eq.
(\ref{dmc_alg}).
Using Eqs. (\ref{def_omega}) and (\ref{dmc_alg}) the expectation value at time
step $i$ becomes
\ben
\label{def_eal}
E_i(\alpha) = \sum_j^{N_w} E^L_{i,j}(\alpha)\prod_{k=1}^{i} e^{-\Delta t
\left(E^L_{k,j}(\alpha)-E^0_i(\alpha)\right)}. \een 
Here, $E^L_{i,j}(\alpha)=E^L_{i,j}+\alpha O^L_{i,j}$ and
$E^0_i(\alpha)= E^0_i+\Delta E^0_i(\alpha)$, so the weight of the wavefunction
is 
\bea \Omega_i&=&\sum_j^{N_w} \underbrace{ \exp\left(-\Delta t
\sum_{k=1}^{i}\left(E^L_{k,j}-E^0_i\right)\right)}_{\omega_{i,j}}
\nonumber\\&& \times\exp\left({-\Delta t \sum_{k=1}^{i}\left(\alpha
O^L_{k,j}-\Delta E^0_i(\alpha)\right)}\right) 
\\ &=&
\overline{
\exp\left(-t\alpha X_{i}\right)}
\exp\left(t\Delta E^0_i(\alpha)\right),
\eea
where $X_{i,j}=\frac{1}{i}\sum_{k=1}^{i} O^L_{k,j}$ and $t=i\Delta t$,
and we have made use of the fact that the growth estimator
$\Delta E^0_i(\alpha)$ is independent of the index $j$. $\Delta E^0_i(\alpha)$
ensures that $\Omega_i(\alpha)=1$ to all orders of $\alpha$, hence
$\Delta E^0_i(\alpha)=
-\frac{1}{t}\log\left[\overline{
\exp\left(-t\alpha X_{i}\right)}\right]$.
Eq. (\ref{def_eal}) then becomes
\ben
\label{def_eal2}
E_i(\alpha)=\frac{
\overline{ E^L_{i}(\alpha)e^{-t\alpha X_{i}}}
}{
\overline{ e^{-t\alpha X_{i}}}
}.
\een
Evaluating
$\Delta E^0_i$ to first order gives the growth estimator of an 
operator:
\ben
\label{def_de0}
O^{GR}_i=\left.\frac{\partial E^0_i(\alpha)}{\partial\alpha}\right|_{\alpha=0}=
\left.\frac{\partial \Delta E^0_i(\alpha)}{\partial\alpha}\right|_{\alpha=0}=
\overline{ X_{i}}.
\een
In other words, the DMC sampling of $X_{i,j}$
by virtue of the HF theorem yields a growth estimator of the 
true expectation value
of $\hat{O}$. 
Interestingly, the growth estimator, if the residual weights are chosen to be
zero, 
appears to be similar to Eq. (13) of Ref.~\cite{no_tagging}.
Applying the HF theorem to the energy estimator 
Eq. (\ref{def_eal2}) yields
a second estimator
\ben
\label{def_de1}
O^{E}_i=\left.\frac{\partial E_i(\alpha)}{\partial\alpha}\right|_{\alpha=0}=
\overline{ O^L_{i} } 
-t\left(
\overline{ E^L_{i} X_{i} }
-\overline{ E^L_{i}} \cdot \overline{ X_{i}}
\right).
\een
Equations (\ref{def_de0}) and (\ref{def_de1})  are of course 
evaluated at $\alpha=0$ and are therefore
accessible in a regular DMC calculation.
We see that for $O^{E}_i$ the standard estimator $\overline{ O^L_{i}}$ is 
augmented by a correction
term $\Delta O^{E}_i=-t\left( \overline{ E^L_{i} X_{i}}
-\overline{ E^L_{i}}\cdot
\overline{ X_{i}} \right)$.  Several observations can be made. First,
in the case of the $\Psi_T$ being the ground state $\Psi^{fn}_0$
for a given nodal structure
the correction term is zero ($ E^L_{i,j}$ is a constant!) 
and only  $\overline{ O^L_{i}}$ contributes as it
should. Furthermore, the new estimator $O^{E}_{i}$ and the usual one 
 $\overline{ O^L_{i}}$
sample an observable and are both independent of the auxiliary
DMC parameter $t$. It follows that $
\overline{ E^L_{i} X_{i}} -\overline{ E^L_{i}}\cdot
\overline{ X_{i}}
\sim \frac{1}{t}$.
Thirdly, since the growth estimator Eq.
(\ref{def_de0}) is derived from the ``averaged'' quantity $E^0_i$ rather than 
$\tilde{E}^0_i$, Eq. (\ref{def_de0}) is itself already averaged over $i$ and therefore
the final estimate at $i$. This is in contrast to  Eq. (\ref{def_de1}) which still has to be 
averaged over all $i$ to yield the final DMC estimate. Using $\tilde{E}^0_i$ yields
an estimator  $\tilde{O}^{GR}_i$ which when averaged over $i$ gives $O^{GR}_i$.
Finally, within the fixed-node approximation the correction term in Eq.
(\ref{def_de1}) can 
be viewed as a direct measure of the error of the trial wavefunction with
respect to a certain operator. 
In the remainder of the paper we will only discuss the direct 
estimator Eq. (\ref{def_de1}). 

An important question is which operators are admissible and can be sampled 
using the  HF estimator Eq. (\ref{def_de1}) or for that matter the growth
estimator Eq. (\ref{def_de0})?
Looking at the
definition of the DMC algorithm one sees that it is based on splitting
the Hamiltonian into a kinetic energy kernel that gives rise to the diffusion
part of the algorithm and a potential energy term that has to be diagonal in
real
space. The diffusion part always being the same it follows that
$\Delta \hat{H}=\alpha \hat{O}$ has to be diagonal in real space too. Using
for example $O^L= T^L=\frac{\hat{T}\Psi_T}{\Psi_T}$ therefore
actually corresponds to sampling the real space many-body operator
given by the function $T^L$, rather than the kinetic energy. 
The result using Eq. (\ref{def_de1}) is $\int
\frac{\hat{T}\Psi_T}{\Psi_T} \left[\Psi^{fn}_0\right]^2 dV$ which in general is
not the
desired expectation value 
$\langle \hat{T}\rangle_{fn}=\int
\frac{\hat{T}\Psi^{fn}_0}{\Psi^{fn}_0} \left[\Psi^{fn}_0\right]^2 dV$.
Nevertheless, $\langle \hat{T}\rangle_{fn}$ is accessible within DMC by
using $\langle \hat{T}\rangle_{fn} =\langle \hat{H}\rangle_{fn} - \langle
\hat{V}\rangle_{fn}$ since the last two quantities can be sampled using
standard DMC and HFS, respectively.

In the following, we give a few examples to demonstrate the applicability of
HFS. We apply the method to sample (i) the density of Helium 
and (ii) the Ewald energy of a homogeneous electron gas with and without
interactions. All data are given in atomic units and we used the
CASINO~\cite{casino} package. The target for the number of walkers was between
200 and 400 and the
residual weights were allowed to fluctuate between $0.5$ and $2$. 
While we did not perform extensive studies it seems the 
algorithm works with and without residual weights.
The only modification to the code consisted of adding
a variable $X$ to each walker, updating $X$  and applying Eq. (\ref{def_de1}).
Other than that we used the code as-is in a standard setup.

\begin{figure}
\centering
\includegraphics[width=0.48\textwidth]{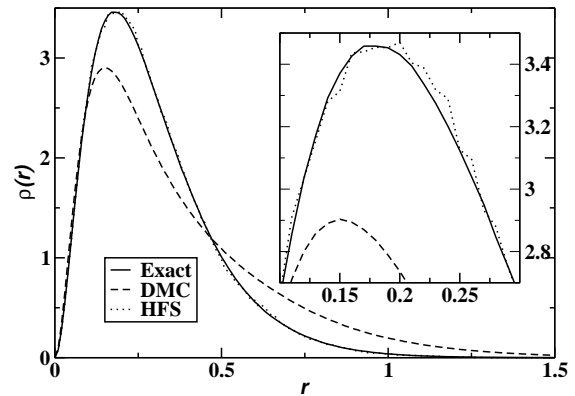}
\caption{
The ``exact'' Helium density (solid line) was derived using the well optimized 
wavefunction provided by the CASINO package: The difference (not shown) between
standard
DMC sampling and HFS is essentially zero ($\Delta O^{E}_i \ll O^{E}_i$). 
In addition, results using
a trial wavefunction with wrong radial function are
presented. Standard DMC yields a smooth but rather poor density. HFS,
while noisier (see inset), follows the correct density
even in the asymptotic region far from the nucleus where despite
little information HFS corrects for the wrong behaviour.}
\label {heden}
\end{figure}

Figure~\ref{heden} shows the electron density (arbitrary units) of He, as
obtained from standard DMC and from our HF method. When the
well-converged (i.e. $\Delta O^{E}_i \ll O^{E}_i$) correlated wavefunction
supplied with CASINO is used both calculations yield essentially the same
result (solid line); when  an ``incorrect'' trial wavefunction (which we have
chosen to be the same as the ``correct'' one but with the radial term heavily
skewed) is used, only our new method (dotted line) recovers the correct
density, albeit the noise in the data is larger. Equally, the interaction
energy is also recovered (DMC correct 
wavefunction: $0.947$, incorrect wavefunction $0.791$, incorrect wavefunction
HFS: $0.958$). We have also performed DMC calculations of the Hydrogen
density, where we systematically deformed the known exact wavefunction. 
Suffice it to say, as for He we again see confirmation of our algorithm. 
An interesting point to add here regards
the extent to which the wavefunction could be skewed. It turns out - rather
plausibly - that if the
wavfunction ceases to actually sample certain parts of phase space HFS cannot
recover
the true form. Nevertheless it seems capable of correcting relatively strong
errors 
in the wavefunction (viz. the rarely sampled asymptoticically decaying part of the wavefunction 
 in Fig. (\ref{heden})),
but the details are clearly a topic for further investigation.

\begin{figure}
\centering
\includegraphics[width=0.48\textwidth]{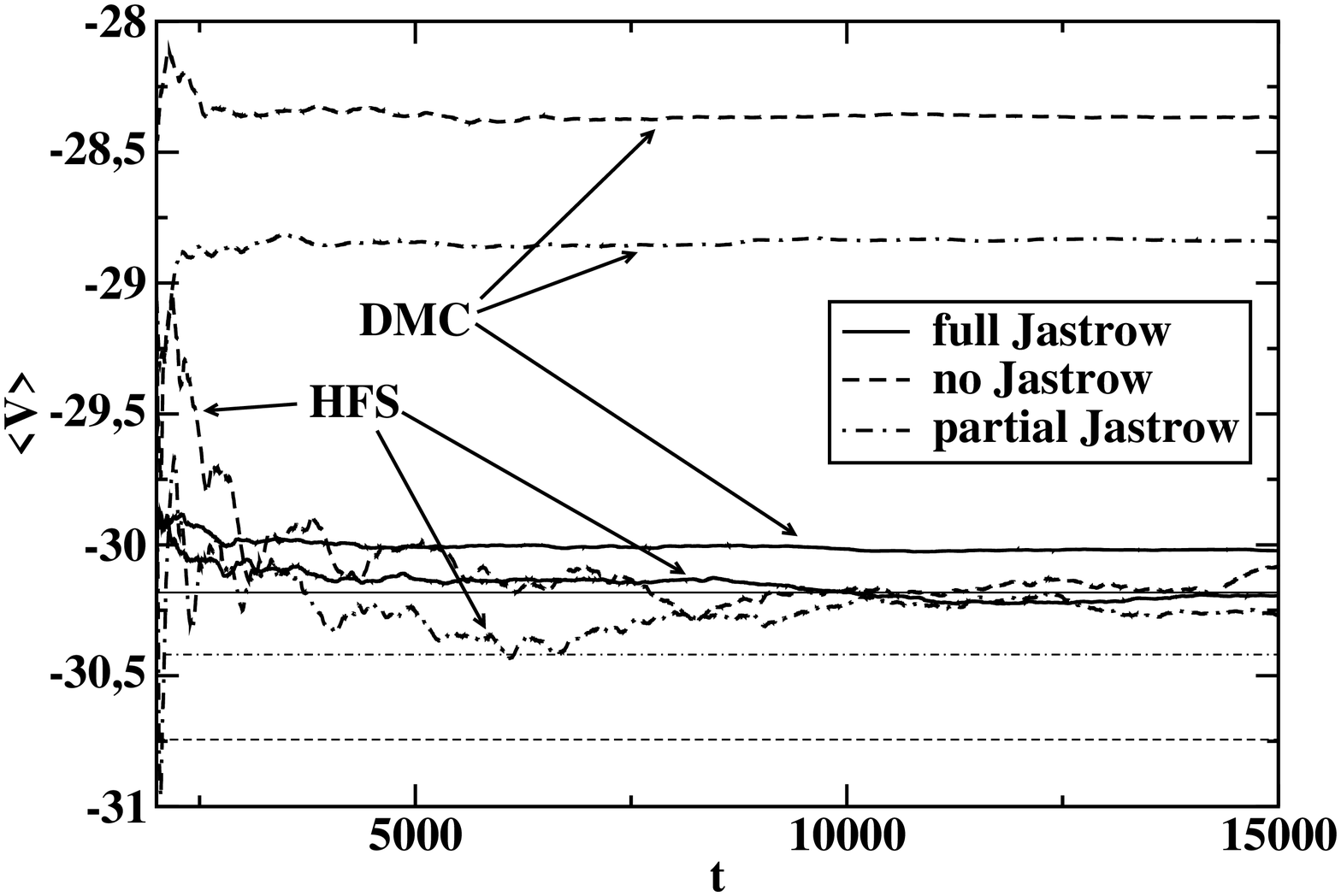}
\caption{
Results for the Ewald energy of an unpolarized homogeneous electron gas
($r_s=1$) with 54 electrons. Standard DMC, $\langle\hat{O}\rangle_{DMC}$,
yields the relatively smooth curves
at the top (see arrows). The noisier data below
use HFS (see arros) and the thin streight lines correspond to $\langle\hat{O}\rangle_{cDMC}$ 
at the end of the run. The partial Jastrow factor contains a correlation term
without cusp.
}
\label {cc=1}
\end{figure}

As in standard DMC sampling, the worse the trial wavefucntion $\Psi_T$,
the larger the noise when using HFS.  
However, when looking at the raw data before averaging over $i$ (not shown) 
we observed that the noise in the HFS data rises
during the progression of the sampling, hence standard error estimation does not work. The source 
can be traced to sampling over histories $X_{i}$. Limiting their depth results
in a constant noise term though also reintroduces a systematic bias.
Also, in a recent paper~\cite{wbias} Warren and Hinde
observe that using the forward-walking
method in DMC necessitates a rapidly growing number of walkers as the
dimensionality
of the quantum system is increased.
These two issues then lead us to the question as to whether HFS works
for larger systems. 
We have therefore looked at an unpolarized  homogeneous electron gas at
$r_s=1$. We used 
a finite simulation cell (periodic boundary condition) with 54 electrons. 
The data we plot shows the Ewald
interaction energy with no additional finite size corrections. We show in
Fig.~\ref{cc=1} results
for a fully interacting system that we have obtained by using trial
wavefunctions with either no Jastrow factor, a
partially optimized Jatrow factor, or a fully optimized one.
 We show the mixed DMC estimate $\langle\hat{O}\rangle_{DMC}$, 
the corrected estimate  
$\langle \hat{O}\rangle_{cDMC},
=2\langle \hat{O}\rangle_{DMC}-\langle
\hat{O}\rangle_{VMC}$
which contains a second-order error,
and the results for HFS. The MC runs
start at 0 with a
short equilibration phase and we start sampling at time step $2000$.
 The corrected estimate using the fully optimized Jastrow factor ought to give
the best result.
Clearly all three HFS estimates are very close but especially the non-optimized 
wavefunctions yield quite noisy data. Nevertheless, even in that case the
results are a lot better 
than using the pure DMC output for the best wavefunction. However, they are all
also better 
than the corrected $\langle\hat{O}\rangle_{cDMC}$ results of
the partially/non-optimized wavefunction. 
Regarding the noise one also has to keep 
in mind the difficulty of the task: The interaction energy is dominated by the
region 
where the electrons get close to each other but that is where the error of the
non-optimized
wavefunctions is largest. HFS essentially has to build a cusp from scratch.

\begin{figure}
\centering
\includegraphics[width=0.48\textwidth]{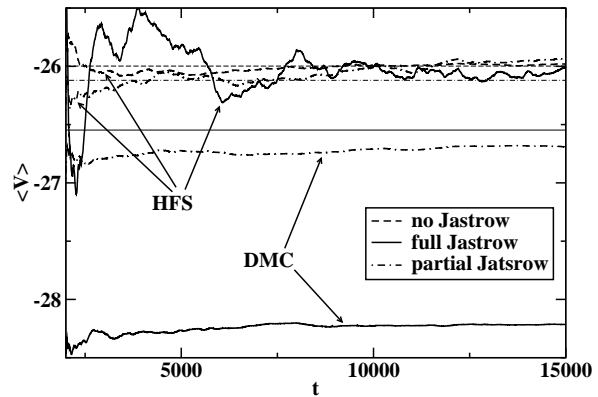}
\caption{
As in Fig.~\ref{cc=1}, but for an interaction-free Hamiltonian.
The noisier data at the top have been sampled using the HFS estimator (see arrows). Below
follow the relatively smooth standard DMC values (see arrows) 
sampling $\langle\hat{O}\rangle_{DMC}$, except for the no-Jastrow
case where the two estimators yield the same data as $\Delta O^{E}_i=0$.
The thin streight lines correspond to $\langle\hat{O}\rangle_{cDMC}$
 at the end of the run, except in the case of 
no Jastrow factor where the thin line gives the essentially exact VMC value.}
\label {cc=0}
\end{figure}

Fig.~\ref{cc=0} repeats the same analysis
for a non-interaction Hamiltonian where the Slater determinant (no Jastrow
factor) 
is the exact solution, whence the 
HFS data and the standard DMC data in that case being identical. 
This is of course consistent with
Eq. (\ref{def_de1}) and proves that given the correct nodes, HFS
yields the 
correct answer. Apart from that Fig.~\ref{cc=0} is essentially a mirror image
of Fig.~\ref{cc=1}. In general, we see that unless the wavefunction is well
optimized 
the HFS estimate is considerably better, despite the noise in the data. Such
situations might 
occur when the system is dominated by the bulk while we are interested in
sampling data in the surface region. Optimization based on the total energy or
variance 
would result in a sub-optimal wavefunction away from the bulk and hence
erroneous standard
sampling.

In conclusion, by applying the HF theorem directly to the DMC algorithm 
we have introduced a new method to sample a large class of operators exactly
within standard DMC.
Our method works for both small and large systems and is easy to add to 
standard DMC, enabling the sampling of a large class of operators (densities,
interaction energies, etc.): Only one extra variable per operator ($X_{i,j}$) 
need be added to the walkers, involving no more than an extra summation step
during sampling; 
simple algebra (Eqs. (\ref{def_de0}) and (\ref{def_de1})) does the rest.
Future work
is needed to better understand, estimate, and deal with the noise and its slow increase
with simulation time. 
This is currently under investigation. Similarly, the effect of residual
weights needs to be 
looked at in more detail. 
 A promising line of research already under way
is to look at the second derivative. This might allow efficient DMC sampling of
the 
fixed-node 
density-response function and related quantities, the study of which is
currently not feasible due
to being numerically too demanding.
       
R.G. would like to thank W.M.C. Foulkes, B. Wood, and N. Hine of Imperial
College 
for helpful discussions.
The authors acknowledge partial support by the University of the
Basque Country, the Basque Unibertsitate eta Ikerketa Saila, the MCyT, and the
EC 6th framework Network of Excellence NANOQUANTA (Grant No.
NMP4-CT-2004-500198).


\begin{thebibliography}{100}
\bibitem{HFproblem} When using the HF theorem in conjunction with numerical
derivatives, problems appear if the underlying Hilbert space changes
with $\alpha$. This is not the case here.
\end{thebibliography}
\end{document}